\documentclass[a4paper, 12pt]{article}
\usepackage{amsmath}
\numberwithin{equation}{section}
\begin{document}
\begin{flushright}
    hep-th/0607065 \\
    DAMTP-2006-56
  \end{flushright}
\vskip 1cm
  \begin{center}
\LARGE{\textbf{Separability of multi-charge black holes in supergravity}}
\end{center}
\begin{center}
        \begin{center}
        {\Large Paul Davis\footnote{P.Davis@damtp.cam.ac.uk} }\\
        \bigskip\medskip
        {\it  DAMTP, Centre for Mathematical Sciences,\\
      University of Cambridge,
      Wilberforce Rd.,\\
      Cambridge CB3 0WA, UK\\}
        \end{center}
  \end{center}
  \vskip 0.5cm


\begin{abstract}
\noindent
In this paper, we show that some five-dimensional rotating black hole
solutions of both gauged and ungauged supergravity, with independent
rotation parameters and three charges admit separable solutions to the
massless Hamilton-Jacobi and Klein-Gordon equations. This allows us to
write down a conformal Killing tensor for the spacetime. Conformal
Killing tensors obey an equation involving a co-vector field. We find
this co-vector field in three specific examples, and also give a
general formula for it.   
\end{abstract} 


\section{Introduction}
It has long been a goal of theoretical physics to understand the
origin of the Bekenstein-Hawking entropy of black holes. One
potentially fruitful way of attacking this problem is by using
holography and, in particular, the AdS/CFT correspondence
\cite{maldacena}. From this perspective, it is useful to study
asymptotically AdS solutions of gauged supergravity since this theory 
arises after reducing ten or eleven dimensional supergravity theories
on an $S^n$. Type IIB supergravity can be reduced on an $S^5$ to give
five dimensional gauged supergravity in an asymptotically $AdS_5$
spacetime. In this case, the boundary theory will then be $\mathcal{N}
= 4$ super-Yang-Mills theory which is relatively well
understood. Eleven-dimensional supergravity can be reduced on an
$S^7$, but then the dual theory lives on the boundary of $AdS_4$,
and is poorly understood. In the $D=5$ case, the $S^5$ has isometry
group $SO(6)$ which contains $U(1)^3$ as a Cartan subgroup. In this
paper, we study the hidden symmetries of black hole solutions that are
charged under the three $U(1)$s.      

\vskip 0.2cm
\noindent
Higher dimensional black holes are currently an area of considerable
interest. Higher dimensional, asymptotically flat, Kerr black holes
were first found in \cite{mp} and the asymptotically (A)dS analogues
were found more recently in \cite{glpp}. The five-dimensional
Myers-Perry and Kerr-(A)dS black holes were shown to possess Killing
tensors for arbitrary rotation parameters in  \cite{stoj} and \cite{kl1}
respectively, and in the case where the rotation parameters can take
one of two different values, a Killing tensor has been found in all
dimensions \cite{vs}.  

\vskip 0.2cm
\noindent
Generalisations to rotating black holes with electrical charge have
been found. In the case of $D=5$ minimal gauged supergravity, an
electrically charged black hole with equal rotation parameters was
found in \cite{clp4} and was shown to possess a reducible Killing
tensor in \cite{kl2}. The solution with unequal rotation parameters
was found in \cite{cclp2} and an irreducible Killing tensor for this
solution was computed in \cite{dkl}.

\vskip 0.2cm
\noindent
Other generalisations have also been discovered. For example, higher
dimensional, black holes with NUT charge and a single, non-zero, rotation
parameter were found in \cite{klemm}, and Killing tensors for these
solutions were found in \cite{cglp}. A
class of cohomogeneity-2, rotating, AdS black holes in higher
dimensions with a generalisation of the NUT parameter was found in
\cite{clp} and the Killing tensors were computed in \cite{clp2,
  davis}. Kerr-NUT-AdS black holes with arbitrary rotation parameters
have recently been found in \cite{clp3}.

\vskip 0.2cm
\noindent
In this paper, we investigate the separability of some
five-dimensional black hole solutions with three charges and two
arbitrary rotation parameters. The first black hole solution that we
study is the general solution in ungauged $\mathcal{N}=2$ supergravity
with three unequal charges that was found in \cite{cy}. The second
solution arises from $SO(6)$ gauged supergravity and has two of its
three charges equal \cite{cclp}. The final black hole solution that we
study is the supersymmetric, $U(1)^3$ black hole found in \cite{klr}.

\vskip 0.2cm
\noindent
All the metrics that we work with have cohomogeneity-2, and the
inverse metric has the property that when multiplied by some scalar
function, it separates additively into tensors depending on only the
two non-trivial coordinates. Unfortunately, the function that we have
to multiply the inverse metric by is complicated, so we can only
separate the Hamilton-Jacobi equation in the case of null
geodesics. This introduces a constant of the motion that is quadratic
in the canonical momenta and is due to the existence of a non-trivial
conformal Killing tensor. The fact that all of these black hole
solutions possess a conformal Killing tensor is unsurprising as they
can all be put into the canonical form presented in \cite{cclp3}.

\vskip 0.2cm
\noindent
Several examples of such black holes are known, and in this paper, we
investigate whether they are geodesically integrable. In fact we find
that this is only possible if we consider massless particles. This is
related to the existence of a non-trivial conformal Killing tensor
which obeys the equation $\nabla_{( \mu} K_{\nu \sigma)} = g_{( \mu
\nu} V_{\sigma)}$ for some co-vector field $V_{\sigma}$. In this paper
we find a general formula for the co-vector field based on fairly
relaxed conditions on the spacetime metric which all known
three-charge black holes satisfy.

\vskip 0.2cm
\noindent
There is some ambiguity in the way in which we separate the
Hamilton-Jacobi equation. For the cohomogeneity-2 metrics that we are
considering in this paper, the massless Hamilton-Jacobi equation gives
a constant of the motion that is quadratic in the canonical momenta,
and can be written in two different ways. Thus, the separation of the
Hamilton-Jacobi equation gives rise to two distinct conformal Killing
tensors. It turns out that these generate two different co-vectors,
which differ by the gradient of a scalar function This can be traced
back to the fact that the two Killing tensors differ by a scalar
function multiplying the metric. It seems that these co-vector fields
act rather like $U(1)$ gauge fields and an antisymmetric field
strength tensor can be defined that is invariant under such gauge
transformations and perhaps it is this that should be regarded as the
entity with physical relevance. 

\vskip 0.2cm
\noindent
The structure of this paper is as follows. In section 2, we consider
the three charge solution of ungauged supergravity found in
\cite{cy}. We show that the Hamilton-Jacobi equation is separable in
the massless case and we show that this is due to the existence of a
non-trivial conformal Killing tensor which we find. We then compute
the associated co-vector field and show that the Klein-Gordon equation
is also separable in the massless case. In section 3, we repeat this
analysis for the non-extremal black hole solution of $U(1)^3$ gauged
supergravity with two charges set equal and the third related to it
\cite{cclp}. In section 4, we show that the same results hold true for
the recently found supersymmetric solution of $U(1)^3$ gauged
supergravity found in \cite{klr}. In section 5 we prove some general
results about the form of the associated co-vector fields and show
that they control the evolution of a scalar function constructed from
the conformal Killing tensor along the worldline of a massive
particle. Finally, in section 6 we present our conclusions.  


\section{Three-charge black holes from ungauged supergravity}
In this section, we examine the three-charge five-dimensional black
holes found in \cite{cy}. In ungauged supergravity, one can use
solution generating techniques to add charges to the neutral
Myers-Perry black hole \cite{mp}. The solution that was found in
\cite{cy} can be written in the following form:  
\begin{eqnarray}
d s^2 &=& \bar{\Delta}^{\frac{1}{3}} \Bigg[ - \frac{\rho^2 (\rho^2 - 2
    m)}{\bar{\Delta}} d t^2 + \frac{ d r^2}{\Delta_r(r)} + d \theta^2
    \nonumber \\
&-& \frac{4 m \sin^2 \theta}{\bar{\Delta}} \Big( \rho^2 ( a \,
    \mathrm{ch}_e  \, \mathrm{ch}_{e1} \, \mathrm{ch}_{e2} - b \,
    \mathrm{sh}_e \, \mathrm{sh}_{e1} \, \mathrm{sh}_{e2} ) + 2 m b \,
    \mathrm{sh}_e \, \mathrm{sh}_{e1} \, \mathrm{sh}_{e2} \Big) d t d
    \phi \nonumber \\ 
&-& \frac{4 m \cos^2 \theta}{\bar{\Delta}} \Big( \rho^2 ( b \,
    \mathrm{ch}_e \, \mathrm{ch}_{e1} \, \mathrm{ch}_{e2}  - a \,
    \mathrm{sh}_e \, \mathrm{sh}_{e1} \, \mathrm{sh}_{e2} ) +2 m a \,
    \mathrm{sh}_e \, \mathrm{sh}_{e1} \, \mathrm{sh}_{e2} \Big) d t d
    \psi \nonumber \\  
&+& \frac{\sin^2 \theta}{\bar{\Delta}} \Big(  (r^2 + a^2 + 2 m \,
    \mathrm{sh}^2_e) (\rho^2 + 2 m \, \mathrm{sh}^2_{e1}) ( \rho^2 + 2
    m \, \mathrm{sh}^2_{e2} ) \nonumber \\
&+&2 m \sin^2 \theta \{ ( a^2 \, \mathrm{ch}^2_e - b^2 \,
    \mathrm{sh}^2_e)  \rho^2 + 4 m \, a \, b  \, \mathrm{sh}_e \,
    \mathrm{sh}_{e1}  \, \mathrm{sh}_{e2} \, \mathrm{ch}_e \,
    \mathrm{ch}_{e1} \, \mathrm{ch}_{e2} \nonumber \\
&-& 2 m \, \mathrm{sh}^2_{e1} \, \mathrm{sh}^2_{e2} (
    a^2 \, \mathrm{ch}^2_e + b^2 \, \mathrm{sh}^2_e) - 2 m b^2 \,
    \mathrm{sh}^2_e ( \, \mathrm{sh}^2_{e1} + \, \mathrm{sh}^2_{e2})
    \} \Big) d \phi^2 \nonumber \\ 
&+& \frac{\cos^2 \theta}{\bar{\Delta}} \Big(  (r^2 + b^2 + 2 m \,
    \mathrm{sh}^2_e) (\rho^2 + 2 m \, \mathrm{sh}^2_{e1}) ( \rho^2 + 2
    m \, \mathrm{sh}^2_{e2} ) \nonumber \\
&+& 2 m  \cos^2 \theta \{ ( b^2 \, \mathrm{ch}^2_e - a^2 \,
    \mathrm{sh}^2_e) \rho^2  + 4 m \, a \, b  \, \mathrm{sh}_e \,
    \mathrm{sh}_{e1} \, \mathrm{sh}_{e2} \, \mathrm{ch}_e \,
    \mathrm{ch}_{e1} \, \mathrm{ch}_{e2} \nonumber \\ 
&-& 2 m \, \mathrm{sh}^2_{e1} \, \mathrm{sh}^2_{e2} (
    b^2 \, \mathrm{ch}^2_e + a^2 \, \mathrm{sh}^2_e) -2 m a^2 \,
    \mathrm{sh}^2_e ( \, \mathrm{sh}^2_{e1} + \, \mathrm{sh}^2_{e2})
    \} \Big) d \psi^2 \nonumber \\  
&+& \frac{4 m \sin^2 \theta \cos^2 \theta}{\bar{\Delta}} \Big( a b
    \{ \rho^2 - 2 m ( \mathrm{sh}^2_{e1} \, \mathrm{sh}^2_{e2} + \,
    \mathrm{sh}^2_e \, \mathrm{sh}^2_{e1} + \mathrm{sh}^2_e \,
    \mathrm{sh}^2_{e2}) \} \nonumber \\ 
&+& 2 m \{ (a^2 + b^2) \, \mathrm{ch}_e \, \mathrm{ch}_{e1} \,
    \mathrm{ch}_{e2} \, \mathrm{sh}_e \, \mathrm{sh}_{e1} \,
    \mathrm{sh}_{e2} - 2 a b  \, \mathrm{sh}^2_e \, \mathrm{sh}^2_{e1} \,
    \mathrm{sh}^2_{e2} \} \Big) d \phi d \psi \Bigg]. 
\end{eqnarray}
Here, we have used the notation $\mathrm{sh}_e \equiv \sinh
\delta_e$, $\mathrm{sh}_{e1} \equiv \sinh \delta_{e1}$ and
$\mathrm{sh}_{e2} \equiv \sinh \delta_{e2}$ where $\delta_e$,
$\delta_{e1}$ and $\delta_{e2}$ are the $SO(1,1)$ boosts that were
applied to the neutral rotating solution in order to obtain the three
charge solution. Also, $\mathrm{ch}_e \equiv \cosh \delta_e$ and
likewise for the other boosts. We have defined the following functions:
\begin{equation}
\bar{\Delta} = (\rho^2 + 2 m \, \mathrm{sh}^2_{e}) (\rho^2 + 2
m \, \mathrm{sh}^2_{e1}) (\rho^2 + 2 m \, \mathrm{sh}^2_{e2}), 
\end{equation}
\begin{equation}
\rho^2 = r^2 + a^2 \cos^2 \theta + b^2 \sin^2 \theta,
\end{equation}
and
\begin{equation}
\Delta_r(r) = \frac{ (r^2+ a^2)(r^2 + b^2) - 2 m r^2}{r^2}.
\end{equation}
 
\vskip 0.2cm
\noindent
The gauge fields that give rise to the three charges have the
following origin: The first one is the Kaluza-Klein gauge field
associated with the first compactified direction, the second gauge
field comes from the ten-dimensional two form field associated to the
first compactified direction, and the field strength of the third
gauge field is dual to $H_{\mu \nu \rho}$ which is the three-form
field strength of the five-dimensional two-form $B_{\mu \nu}$. The
duality between the three and two-form field strengths is given by
\begin{equation}
H^{\mu \nu \rho} = - \frac{e^{4 \varphi/3}}{2! \sqrt{-g}} \varepsilon^{\mu
  \nu \rho \lambda \sigma} F_{\lambda \sigma},
\end{equation}
where the dilaton field, $\varphi$ is given by the equation
\begin{equation}
e^{2 \varphi} = \frac{(\rho^2 + 2 m \mathrm{sh}^2_e)^3}{\bar{\Delta}}. 
\end{equation}

\vskip 0.2cm
\noindent
The determinant of this metric is surprisingly simple. We find that 
\begin{equation}
\sqrt{-\mathrm{det} g} =  r  \sin \theta \cos \theta \bar{\Delta}^{1/3}.  
\end{equation}

\vskip 0.2cm
\noindent
The inverse metric is also relatively simple and has the non-trivial
property that $\bar{\Delta}^{1/3} \, g^{\mu \nu}$ is additively
separable as a function of $r$ and a function of $\theta$. We obtain:  
\begin{eqnarray}
\bar{\Delta}^{1/3} \, g^{t t} &=& - a^2 \cos^2 \theta - b^2 \sin^2
\theta - r^2 - 2 m ( \mathrm{sh}^2_e +  \mathrm{sh}^2_{e1} +
\mathrm{sh}^2_{e2}) - 2 m  \nonumber \\ 
&-& \frac{4 m^2}{r^2 \Delta_r} \Big( r^2 (1 + \mathrm{sh}^2_e
+ \mathrm{sh}^2_{e1} + \mathrm{sh}^2_{e2} +
\mathrm{sh}^2_e \, \mathrm{sh}^2_{e1} +   \mathrm{sh}^2_e \,
\mathrm{sh}^2_{e2}   \nonumber \\  
&+& \mathrm{sh}^2_{e1} \, \mathrm{sh}^2_{e2}) - (a^2 + b^2) \,
\mathrm{sh}^2_e \, \mathrm{sh}^2_{e1} \, \mathrm{sh}^2_{e2} \nonumber
\\  
&+& 2 a b \mathrm{ch}_e \, \mathrm{ch}_{e1} \, \mathrm{ch}_{e2}
\mathrm{sh}_e \,  \mathrm{sh}_{e1} \mathrm{sh}_{e2} \Big), \nonumber \\
\bar{\Delta}^{1/3} \, g^{t \phi} &=& -\frac{2 m}{r^2 \Delta_r} \Bigg(
  r^2(a \, \mathrm{ch}_e \, \mathrm{ch}_{e1} \,
  \mathrm{ch}_{e2} - b \, \mathrm{sh}_e \, \mathrm{sh}_{e1} \,
  \mathrm{sh}_{e2} ) \nonumber \\
&+&  b (2 m - b^2) \, \mathrm{sh}_e \, \mathrm{sh}_{e1}
  \, \mathrm{sh}_{e2} + a b^2 \, \mathrm{ch}_e \, \mathrm{ch}_{e1} \,
  \mathrm{ch}_{e2} \Bigg), \nonumber
  \\ 
\bar{\Delta}^{1/3} \, g^{t \psi} &=& - \frac{2 m}{r^2 \Delta_r} \Bigg(
  r^2 ( b \, \mathrm{ch}_e \, \mathrm{ch}_{e1} \,
  \mathrm{ch}_{e2} - a \, \mathrm{sh}_e \, \mathrm{sh}_{e1} \,
  \mathrm{sh}_{e2}  ) \nonumber \\
&+&  a (2 m - a^2) \, \mathrm{sh}_e \, \mathrm{sh}_{e1}
  \, \mathrm{sh}_{e2} +a^2 b \, \mathrm{ch}_e \, \mathrm{ch}_{e1} \,
  \mathrm{ch}_{e2} \Bigg),
  \nonumber \\ 
\bar{\Delta}^{1/3} \, g^{\phi \phi} &=& \frac{1}{\sin^2 \theta} -
\frac{2 m b^2 + (r^2 + b^2)(a^2 - b^2)}{r^2 \Delta_r}, \nonumber 
\end{eqnarray}
\begin{eqnarray}
\bar{\Delta}^{1/3} \, g^{\psi \psi} &=& \frac{1}{\cos^2 \theta} -
\frac{2 m a^2 + (r^2 + a^2)(b^2 - a^2)}{r^2 \Delta_r}, \nonumber \\
\bar{\Delta}^{1/3} \, g^{\phi \psi} &=& - \frac{2 m a b}{r^2
  \Delta_r}, \nonumber \\
\bar{\Delta}^{1/3} \, g^{r r} &=& \Delta_r, \nonumber \\
\bar{\Delta}^{1/3} \, g^{\theta \theta} &=& 1.
\end{eqnarray}

\subsection{The Hamilton-Jacobi equation}
The Hamilton-Jacobi equation can be used to generate the equations of
the geodesics in this background. If the Hamilton-Jacobi equation is
completely separable, then geodesic motion will be integrable. For free,
uncharged particles, the Hamiltonian is $H = \frac{1}{2} g^{\mu \nu}
p_{\mu} p_{\nu}$, where $\{p_{\mu} \}$ are the canonical momenta, and
the corresponding Hamilton-Jacobi equation is 
\begin{equation} \label{hjeqn}
\frac{\partial S}{\partial \tau} + \frac{1}{2} g^{\mu \nu}
\frac{\partial S}{\partial x^{\mu}} \frac{\partial S}{\partial
  x^{\nu}} = 0,
\end{equation}
where $S$ is Hamilton's principal function and $\tau$ is the parameter
along the worldline of the particle. Due to the isometries that are
present in this background, $t$, $\phi$ and $\psi$ all appear linearly
in S, so we have:
\begin{equation} \label{hjansatz}
S = \frac{1}{2} m^2 \tau - E t + \Phi \phi + \Psi \psi + F(r, \theta).
\end{equation}
If we multiply (\ref{hjeqn}) by $\bar{\Delta}^{\frac{1}{3}}$, then we
can use the additive separability of $\bar{\Delta}^{\frac{1}{3}}
g^{\mu \nu}$ to try to fully separate $S$. Substituting
(\ref{hjansatz}) into (\ref{hjeqn}) and multiplying through by
$2 \bar{\Delta}^{\frac{1}{3}}$, we obtain
\begin{equation}
m^2 \bar{\Delta}^{\frac{1}{3}} + \Big( \bar{\Delta}^{\frac{1}{3}}
g^{\mu \nu} \Big) \frac{\partial S}{\partial x^{\mu}} \frac{\partial
  S}{\partial x^{\nu}} = 0,
\end{equation}
which will not allow us to separate the $r$ and $\theta$ dependence of
$F(r, \theta)$ for arbitrary rotation parameters and charges, due to
the complicated form of 
$\bar{\Delta}^{\frac{1}{3}}$, unless $m^2 =0$. In the much less
general cases where the rotation parameters are equal, $a=b$, or if
the charges are all equal, $\delta_{e} = \delta_{e1} = \delta_{e2}$,
the Hamilton-Jacobi equation will separate for arbitrary $m^2$.
Neither of these cases are of interest to us, so we deduce that the 
Hamilton-Jacobi equation with arbitrary rotation parameters and
arbitrary charges will only separate if the mass, $m$, is set to
zero. With $m$ set to zero, we can write $F(r,\theta) = S_r(r) +
S_{\theta}(\theta)$ and then the $\theta$-dependent  part of the separated
Hamilton-Jacobi equation is 
\begin{equation}
K = - (a^2 \cos^2 \theta + b^2 \sin^2 \theta) E^2 +
\frac{\Phi^2}{\sin^2 \theta} + \frac{\Psi^2}{\cos^2 \theta} + \Bigg(
\frac{ d S_{\theta}}{d \theta} \Bigg)^2,
\end{equation}
where $K$ is the separation constant. By writing $K = K^{\mu \nu} p_{\mu}
p_{\nu}$, where $p_{\mu}$ is the canonical momentum, we find the
following expression for the symmetric rank 2 tensor, $K^{\mu \nu}$:
\begin{equation}
K^{\mu \nu} = - (a^2 \cos^2 \theta + b^2 \sin^2 \theta) \delta_t^{\mu}
\delta_t^{\nu} + \frac{\delta_{\phi}^{\mu} \delta_{\phi}^{\nu}
}{\sin^2 \theta}  + \frac{\delta_{\psi}^{\mu}
  \delta_{\psi}^{\nu}}{\cos^2 \theta} + \delta_{\theta}^{\mu}
\delta_{\theta}^{\nu}.
\end{equation}
It is easy to check that $K^{\mu \nu}$ is a conformal Killing tensor,
obeying the equation
\begin{equation} \label{ckt}
\nabla_{(\mu} K_{\nu \rho)} = g_{(\mu \nu} V_{\rho)},
\end{equation}
where $V_{\mu}$ is some co-vector field associated to the conformal
Killing tensor. By contracting indices in equation (\ref{ckt}), one
obtains the following equation for $V^{\mu}$,
\begin{equation} \label{cktvec}
V^{\mu} = \frac{2}{7} \nabla_{\nu} K^{\mu \nu} + \frac{1}{7}
\nabla^{\mu} K^{\nu}_{\phantom{\nu} \nu},
\end{equation}
which gives us the very simple expressions for the (co-)vector field, 
\begin{equation}
V^{\mu} =  \frac{\partial_{\theta}
  \bar{\Delta}^{\frac{1}{3}}}{\bar{\Delta}^{\frac{1}{3}}}
  \delta_{\theta}^{\mu} \Longleftrightarrow V_{\mu} =  \Big(
  \partial_{\theta} \bar{\Delta}^{\frac{1}{3}} \Big)
  \delta^{\theta}_{\mu}. 
\end{equation}


\subsection{The Klein-Gordon equation}
In the same way that we can separate the Hamilton-Jacobi equation in
the case where the mass vanishes, we can separate the Klein-Gordon
equation in the case of massless particles. This is because we would
have to multiply the mass term, $M^2 \Phi$, by $\bar{\Delta}^{1/3}$,
which is a complicated function of $r$ and $\theta$. 

\vskip 0.2cm
\noindent
The massless Klein-Gordon equation can be written in the following
form:
\begin{equation} \label{masslesskg}
\partial_{\mu} \Big( \sqrt{-g} \, g^{\mu \nu} \partial_{\nu} \Phi \Big) =
0. 
\end{equation}
If we seek a separable solution to this equation, of the form, 
\begin{equation} \label{kgansatz}
\Phi = e^{- i \omega t} e^{i \alpha \phi} e^{i \beta \psi}
\Theta(\theta) R(r),
\end{equation}
then we obtain the following equation for $\Theta$:
\begin{equation}
\frac{1}{\Theta \sin \theta \cos \theta} \frac{d}{d \theta} \Bigg(
\sin \theta \cos \theta \frac{d \Theta}{d \theta} \Bigg) -
\frac{\alpha^2}{\sin^2 \theta} - \frac{\beta^2}{\cos^2 \theta} +
\omega^2 (a^2 \cos^2 \theta + b^2 \sin^2 \theta) = k,
\end{equation}
where $k$ is the separation constant. By making the change of
variable, $z = \sin^2 \theta$, we can rewrite this equation in the
following form:
\begin{eqnarray}
0 &=& \frac{d^2 \Theta}{d z^2} + \Bigg( \frac{1}{z} + \frac{1}{z-1} \Bigg)
\frac{d  \Theta}{d z} - \frac{1}{4 z(1-z)} \Bigg( \frac{\alpha^2}{z} +
\frac{\beta^2}{1-z} \Bigg) \Theta  \nonumber \\
&+& \frac{\omega^2}{4}
\Bigg(\frac{a^2}{z} + \frac{b^2}{1-z} \Bigg) \Theta  -  \frac{k
  \Theta}{4 z(1-z)}. 
\end{eqnarray}
In general, this differential equation has regular singular points at
$z=0$ and $z=1$, and an irregular singular point at $z=\infty$ and it
can be recast in the form of a confluent Heun equation. In the special
case where the rotation parameters are equal, $a=b$, then $z=\infty$
is a regular singular point, and the solution which is regular at
$z=0$ is 
\begin{equation} 
\Theta(\theta) = A \cos^{\beta/2} \theta \sin^{|\alpha|/2} \theta F
\Big( \gamma_+ ,\gamma_- ; 1 + |\alpha|; \sin^2 \theta \Big),
\end{equation}
where
\begin{equation}
\gamma_{\pm} = \frac{ |\alpha| + \beta +1}{2} \pm \frac{1}{2} \sqrt{ 1
  - k + a^2 \omega^2} .
\end{equation}
The corresponding equation for $R(r)$ is complicated and we shall
not examine it here.


\section{Non-extremal black holes in $D=5$, $SO(6)$ gauged supergravity}
In this section, we consider non-extremal, rotating, black hole
solutions of five-dimensional $SO(6)$ gauged supergravity. The
solutions that we shall consider have arbitrary rotation parameters,
giving the metric cohomogeneity-2, and are charged under all three
gauge fields in the $U(1)^3$ subgroup of $SO(6)$. The charges are not
all independent\footnote{In fact, two of the charges are equal.} as
they depend only on a single parameter, $\delta$. The metric to be
considered is \cite{cclp}: 
\begin{eqnarray}
d s^2 &=& H^{-\frac{4}{3}} \Bigg[ - \frac{X(r)}{\rho^2} \Big( d t - a
  \sin^2 \theta \frac{d \phi}{\Xi_a} - b \cos^2 \theta \frac{d \psi}{\Xi_b}
  \Big)^2 \nonumber \\
&+& \frac{C}{\rho^2} \Big( \frac{a b}{f_3} d t - \frac{b \sin^2
  \theta}{f_2} \frac{d \phi}{\Xi_a} - \frac{a \cos^2 \theta}{f_1}
  \frac{d \psi}{\Xi_b} \Big)^2 \nonumber \\
&+& \frac{Z \sin^2 \theta}{\rho^2} \Big( \frac{a}{f_3} d t -
\frac{1}{f_2} \frac{d \phi}{\Xi_a} \Big)^2 + \frac{W \cos^2
  \theta}{\rho^2} \Big( \frac{b}{f_3} d t - \frac{1}{f_1} \frac{d
  \psi}{\Xi_b} \Big)^2  \Bigg] \nonumber \\
&+& H^{\frac{2}{3}} \Big( \frac{\rho^2}{X} d r^2 +
\frac{\rho^2}{\Delta_{\theta}} d \theta^2 \Big),
\end{eqnarray}
where
\begin{equation}
f_1 = r^2 + a^2, \quad f_2 = r^2 + b^2, \quad f_3 = f_1
f_2 + 2 m r^2 \sinh^2 \delta, 
\end{equation}
\begin{equation}
\rho^2 = r^2 + a^2 \cos^2 \theta + b^2 \sin^2 \theta, \quad
\tilde{\rho}^2 = \rho^2 +  2 m \sinh^2 \delta, \quad H=
\frac{\tilde{\rho}^2}{\rho^2} ,
\end{equation}
\begin{equation}
X(r) = \frac{f_1 f_2}{r^2} - 2 m + g^2 (f_1 + 2 m
\sinh^2 \delta)(f_2 + 2 m \sinh^2 \delta) , 
\end{equation}
\begin{equation}
C = f_1 f_2 \Bigg( X + 2 m - \frac{4 m^2 \sinh^4
  \delta}{\rho^2} \Bigg) 
\end{equation}
\begin{equation}
Z = - b^2 C + \frac{f_2 f_3}{r^2} \Big[ f_3 - g^2 r^2 (a^2 - b^2)(
  f_1 + 2 m \sinh^2 \delta)\cos^2 \theta \Big],
\end{equation}
\begin{equation}
W = - a^2 C + \frac{f_1 f_3}{r^2} \Big[ f_3 + g^2 r^2 (a^2 - b^2)(
  f_2 + 2 m \sinh^2 \delta)\sin^2 \theta \Big],
\end{equation}
\begin{equation}
\Xi_a = 1 - a^2 g^2 \quad \mathrm{and} \quad \Xi_b = 1 - b^2 g^2.
\end{equation}

\vskip 0.2cm
\noindent
The three gauge fields are 
\begin{equation}
A^1 = A^2 = \frac{2 m \sinh \delta \, \cosh  \delta}{\tilde{\rho}^2}
\Big( d t - a \sin^2 \theta \frac{d \phi}{\Xi_a} - b \cos^2 \theta
\frac{d \psi}{\Xi_b} \Big) 
\end{equation}
and 
\begin{equation}
A^3 = \frac{2 m \sinh^2 \delta}{\rho^2}
\Big(b \sin^2 \theta \frac{d \phi}{\Xi_a} +  a \cos^2 \theta
\frac{d \psi}{\Xi_b} \Big).
\end{equation}
There are also three scalar fields: $X_1 = X_2 = H^{- \frac{1}{3}}$
and $X_3 = H^{\frac{2}{3}}$. These scalars are not directly relevant
to the study of symmetries on the phase space, but they are important
when computing thermodynamic quantities like charge. 

\vskip 0.2cm
\noindent
The determinant of this metric can readily be computed and we find
that,
\begin{equation}
\sqrt{-\mathrm{det} g} =  \frac{r H^{\frac{2}{3}} \rho^2 \sin \theta
  \cos \theta }{\Xi_a \Xi_b}. 
\end{equation}

\vskip 0.2cm
\noindent
The inverse metric has the remarkable property that $\rho^2
H^{\frac{2}{3}} g^{\mu \nu}$ is additively separable as a function of
$r$ and $\theta$. We find that it is given by
\begin{eqnarray}
\rho^2 H^{\frac{2}{3}} g^{t t} &=& -\frac{ \Xi_a \Xi_b (a^2 \cos^2
  \theta  +b^2 \sin^2 \theta)}{\Delta_{\theta}} + \frac{4 a^2 b^2 g^2
  m^2}{r^2 X} \sinh^4 \delta \nonumber \\ 
&-& \frac{2 m}{X} \Big( a^2
  + b^2 - a^2 b^2 g^2 + \sinh^2 \delta (a^2 + b^2 - 2 a^2 b^2 g^2) \Big) 
\nonumber \\
&-& \frac{2 m a^2 b^2}{r^2 X} \Big( 1 + 2 \sinh^2 \delta - (a^2 + b^2) g^2
  \sinh^2 \delta \Big) \nonumber \\ 
&-& \frac{ \Xi_a \Xi_b}{X} (f_1 + 2 m \sinh^2 \delta) (f_2 +
  2 m \sinh^2 \delta), \nonumber \\
\rho^2 H^{\frac{2}{3}} g^{t \phi} &=& \frac{a \Xi_a \Xi_b
  }{\Delta_{\theta}} - \frac{a \Xi_a}{r^2 X} \Big( \Xi_b f_3 - 2 m b^2[
  g^2 (f_1 + b^2) \sinh^2 \delta - \cosh^2 \delta +  2 m g^2 \sinh^4
  \delta] \Big), \nonumber \\ 
\rho^2 H^{\frac{2}{3}} g^{t \psi} &=& \frac{b \Xi_a \Xi_b
  }{\Delta_{\theta}} - \frac{b \Xi_b}{r^2 X} \Big( \Xi_a f_3 - 2 m a^2[
  g^2 (f_2 + a^2) \sinh^2 \delta - \cosh^2 \delta + 2 m g^2  \sinh^4
  \delta] \Big), \nonumber \\
\rho^2 H^{\frac{2}{3}} g^{\phi \phi} &=& \frac{ \Xi_a^2 ( \Xi_b +
  \cot^2 \theta) }{\Delta_{\theta}} - \frac{\Xi_a^2}{r^2
  X} \Big( - g^2 b^2 f_3 + (a^2 - b^2) f_2 + 2 m b^2 [ 1 -  g^2 (f_1 +
  b^2) \sinh^2 \delta \nonumber \\  
  &-&  2 m g^2 \sinh^4 \delta] \Big), \nonumber 
\end{eqnarray}
\begin{eqnarray}
\rho^2 H^{\frac{2}{3}} g^{\psi \psi} &=& \frac{ \Xi_b^2 ( \Xi_a +
  \tan^2 \theta) }{\Delta_{\theta}} - \frac{\Xi_b^2}{r^2
  X} \Big( - g^2 a^2 f_3 + (b^2 - a^2) f_1 +  2 m a^2 [ 1 -  g^2 (f_2
  + a^2) \sinh^2 \delta \nonumber \\  
&-& 2 m g^2 \sinh^4 \delta] \Big), \nonumber \\
\rho^2 H^{\frac{2}{3}} g^{\phi \psi} &=& -\frac{a b g^2 \Xi_a
  \Xi_b}{\Delta_{\theta}}  - \frac{a b \Xi_a \Xi_b}{r^2
  X} \Big( - g^2 f_3  + 2 m  [ 1 -  g^2 (r^2 + a^2 + b^2) \sinh^2
  \delta \nonumber \\
&-& 2 m g^2 \sinh^4 \delta] \Big), \nonumber \\
\rho^2 H^{\frac{2}{3}} g^{\theta \theta} &=& \Delta_{\theta},
\nonumber  \\
\rho^2 H^{\frac{2}{3}} g^{r r} &=& X.  
\end{eqnarray}

\subsection{The Hamilton-Jacobi equation}
In order to check whether the Hamilton-Jacobi equation is separable or
not, we substitute (\ref{hjansatz}) into (\ref{hjeqn}) and multiply
through by $\rho^2 H^{\frac{2}{3}}$. As in the previous section, if
$m \neq 0$, we are unable to separate the Hamilton-Jacobi equation. In
the massless case, the $\theta$-dependent part of the separated
Hamilton-Jacobi equation is  
\begin{eqnarray} 
K &=& -\frac{ \Xi_a \Xi_b (a^2 \cos^2 \theta  +b^2 \sin^2
  \theta)}{\Delta_{\theta}} E^2 -  \frac{2 a \Xi_a \Xi_b
  }{\Delta_{\theta}}  E \Phi -  \frac{2 b \Xi_a \Xi_b
  }{\Delta_{\theta}} E \Psi  \nonumber \\
&+& \frac{ \Xi_a^2 ( \Xi_b + \cot^2 \theta) }{\Delta_{\theta}} \Phi^2
  + \frac{ \Xi_b^2 ( \Xi_a + \tan^2 \theta) }{\Delta_{\theta}} \Psi^2
  -   \frac{2a b g^2 \Xi_a \Xi_b}{\Delta_{\theta}} \Phi \Psi +
  \Delta_{\theta} \Bigg( \frac{d S_{\theta}}{d \theta} \Bigg)^2.
\end{eqnarray} 
The $r$-dependent part is much more complicated, so we restrict our
analysis to the simpler case for the moment. 

\vskip 0.2cm
\noindent
By writing $K = K^{\mu \nu} p_{\mu} p_{\nu}$ where $\{ p_{\mu} \}$ are
the canonical momenta, we find:
\begin{eqnarray} 
K^{\mu \nu} &=& -\frac{ \Xi_a \Xi_b (a^2 \cos^2 \theta  +b^2 \sin^2
  \theta)}{\Delta_{\theta}} \delta_t^{\mu} \delta_t^{\nu} +  \frac{a
  \Xi_a \Xi_b 
  }{\Delta_{\theta}}  ( \delta_t^{\mu} \delta_{\phi}^{\nu} +
  \delta_t^{\nu}\delta_{\phi}^{\mu}) + \frac{b \Xi_a \Xi_b
  }{\Delta_{\theta}} ( \delta_t^{\mu} \delta_{\psi}^{\nu} +
  \delta_t^{\nu}\delta_{\psi}^{\mu})  \nonumber \\
&+& \frac{ \Xi_a^2 ( \Xi_b + \cot^2 \theta) }{\Delta_{\theta}}
  \delta_{\phi}^{\mu} \delta_{\phi}^{\nu} + \frac{ \Xi_b^2 ( \Xi_a +
  \tan^2 \theta) }{\Delta_{\theta}} \delta_{\psi}^{\mu} \delta_{\psi}^{\nu}
  -   \frac{a b g^2 \Xi_a \Xi_b}{\Delta_{\theta}} (
  \delta_{\phi}^{\mu} \delta_{\psi}^{\nu}  +
  \delta_{\phi}^{\nu}\delta_{\psi}^{\mu}) \nonumber \\
&+& \Delta_{\theta} \delta_{\theta}^{\mu} \delta_{\theta}^{\nu}.
\end{eqnarray}
This is a conformal Killing tensor and obeys (\ref{ckt}) with the
vector field given by
\begin{equation}
V^{\mu} = \Delta_{\theta}  \frac{\partial_{\theta} (\rho^2
    H^{2/3})}{\rho^2 H^{2/3}} \delta_{\theta}^{\mu}
    \Longleftrightarrow V_{\mu} = \partial_{\theta} (\rho^2
    H^{2/3}) \delta^{\theta}_{\mu}.
\end{equation}


\subsection{The Klein-Gordon equation}
Again, due to the complicated nature of the function $\rho^2
H^{\frac{2}{3}}$, we can only find separable solutions to the
Klein-Gordon equation in the massless case. Substituting the separable
Ansatz (\ref{kgansatz}) into the massless Klein-Gordon equation
(\ref{masslesskg}), we find the following differential equation for
$\Theta(\theta)$:
\begin{equation}
\frac{1}{\Theta \sin \theta \cos \theta} \frac{d}{d \theta} \Bigg(
\sin \theta \cos \theta \Delta_{\theta} \frac{d \Theta}{d \theta}
\Bigg) + \Bigg[\frac{\Xi_a \Xi_b \omega^2 (a^2 \cos^2 \theta + b^2
    \sin^2 \theta) }{\Delta_{\theta}} + \frac{2 \alpha \omega a \Xi_a
    \Xi_b}{\Delta_{\theta}} \nonumber
\end{equation}
\begin{equation}
 + \frac{2 \beta \omega b \Xi_a \Xi_b}{\Delta_{\theta}} -
 \frac{\alpha^2 \Xi_a^2 (\Xi_b + \cot^2 \theta)}{\Delta_{\theta}} +
 \frac{2 \alpha \beta \Xi_a \Xi_b a b g^2}{\Delta_{\theta}}-
 \frac{\beta^2 \Xi_b^2 (\Xi_a + \tan^2 \theta)}{\Delta_{\theta}}
 \Bigg] = k,
\end{equation}
where $k$ is the separation constant. This equation can be made
 simpler by letting $z = \sin^2 \theta$. Doing this, one finds that
\begin{eqnarray}
0 &=& \frac{d^2 \Theta}{d z^2} + \Bigg( \frac{1}{z} + \frac{1}{z-1}
+ \frac{1}{z-d} \Bigg) \frac{d \Theta}{d z} + \Bigg( \frac{ \Xi_a
  \Xi_b \omega^2 a^2}{4 z \Delta_z^2}+  \frac{ \Xi_a
  \Xi_b \omega^2 b^2}{4 (1-z) \Delta_z^2} \nonumber \\
&+& \frac{a \Xi_a \Xi_b \alpha \omega}{2 z (1-z) \Delta_z^2} + \frac{b
  \Xi_a \Xi_b \beta \omega}{2 z (1-z) \Delta_z^2}  - \frac{\alpha^2
  \Xi_a^2 \Xi_b}{4 z (1-z) \Delta_z^2} - \frac{\alpha^2 \Xi_a^2}{4 z^2
  \Delta_z^2}  \nonumber \\
&-& \frac{\beta^2  \Xi_a \Xi_b^2}{4 z (1-z) \Delta_z^2} - \frac{\beta^2
  \Xi_b^2}{4 z^2 \Delta_z^2} + \frac{\alpha \beta \Xi_a \Xi_b a b
  g^2}{2 z (1-z) \Delta_z^2} - \frac{k}{4 z (1-z) \Delta_z} \Bigg)
\Theta,
\end{eqnarray}
where $d = \Xi_a/g^2(b^2 - a^2)$ and $\Delta_z = g^2 (a^2 -
b^2)(z-d)$. This differential equation has four regular singular
points and can straightforwardly be put into the form of Heun's
equation. If $a=b$, we obtain a differential equation with three
regular singular points whose solution can be given in terms of
hypergeometric functions.


\section{Supersymmetric black holes from $D=5$, $U(1)^3$ gauged
  supergravity} 
We consider the most general known supersymmetric solution of $D=5$,
$U(1)^3$ gauged supergravity which was recently found in
\cite{klr}. It has two angular momenta and three charges corresponding
to the three Maxwell fields. Rather than being a five parameter
theory, the absence of closed timelike curves imposes a constraint, so
there are only four free parameters. This leaves us with the following
metric: 
\begin{eqnarray}
d s^2 &=& - H^{-\frac{2}{3}} ( d t  + \omega_{\phi} d \phi +
\omega_{\psi} d \psi)^2 + H^{\frac{1}{3}} h_{m n} d x^m d x^n.
\end{eqnarray}
Here, $H \equiv H_1 H_2 H_3$ where
\begin{equation}
H_I = 1 + \frac{ \sqrt{\Xi_a \Xi_b} \, (1 + g^2 \mu_I) -
  \Delta_{\theta}}{g^2 r^2}.
\end{equation}
We also have the following definitions
\begin{eqnarray}
h_{m n} d x^m d x^n &=& r^2 \Bigg \{ \frac{ d r^2}{\Delta_r} + \frac{d
  \theta^2}{\Delta_{\theta}}  + \frac{ \cos^2 \theta}{\Xi_b^2} \Big[
  \Xi_b + \cos^2 \theta \Big( \rho^2 g^2 + 2(1 + b g)(a + b)g \Big)
  \Big] d \psi^2 \nonumber \\
&+& \frac{ \sin^2 \theta}{\Xi_a^2} \Big[
  \Xi_a + \sin^2 \theta \Big( \rho^2 g^2 + 2(1 + a g)(a + b)g \Big)
  \Big] d \phi^2 \nonumber \\
&+& \frac{2 \sin^2 \theta \cos^2 \theta}{\Xi_a \Xi_b} \Big[\rho^2 g^2 +
  2(a + b)g + (a + b)^2 g^2 \Big] d \psi d \phi \Bigg \},
\end{eqnarray}
\begin{eqnarray}
\Delta_r &=& r^2 [ g^2 r^2 + (1 + a g + b g)^2 ], \quad \quad
\Delta_{\theta} = 1 - a^2 g^2 \cos^2 \theta - b^2 g^2 \sin^2 \theta,
\nonumber \\
\Xi_a &=& 1 - a^2 g^2, \quad \quad \Xi_b = 1 - b^2 g^2, \quad \quad
\rho^2 = r^2 + a^2 \cos^2 \theta + b^2 \sin^2 \theta,
\end{eqnarray}
\begin{eqnarray}
\omega_{\psi} &=& -\frac{g \cos^2 \theta}{r^2 \Xi_b} \Bigg[ \rho^4 +
  (2r_m^2 + b^2) \rho^2 + \frac{1}{2} \Big( \beta_2 - a^2 b^2 + (a^2 -
  b^2)g^{-2} \Big) \Bigg], \nonumber \\
\omega_{\phi} &=& -\frac{g \sin^2 \theta}{r^2 \Xi_a} \Bigg[ \rho^4 +
  (2r_m^2 + a^2) \rho^2 + \frac{1}{2} \Big( \beta_2 - a^2 b^2 - (a^2 -
  b^2)g^{-2} \Big) \Bigg],
\end{eqnarray}
\begin{equation}
r_m^2 = (a + b) g^{-1} +  a b,
\end{equation}
and 
\begin{equation}
\beta_2 = \Xi_a \Xi_b (\mu_1\mu_2 + \mu_2 \mu_3 + \mu_3 \mu_1) -
  \frac{2 \sqrt{\Xi_a \Xi_b} ( 1 - \sqrt{\Xi_a \Xi_b})}{g^2} (\mu_1 +
  \mu_2 + \mu_3) + \frac{3 (1 - \sqrt{\Xi_a \Xi_b})^2}{g^4}.
\end{equation}
It seems as if this solution depends on five parameters: $\mu_1,
\mu_2, \mu_3, a$ and $b$. However as mentioned above, there is a
constraint, so there are only four independent parameters. The
constraint is 
\begin{equation}
\mu_1 + \mu_2 + \mu_3 = \frac{1}{\sqrt{\Xi_a \Xi_b}} \Big[ 2 r_m^2 +
  3 g^{-2} \Big( 1 - \sqrt{\Xi_a \Xi_b} \Big) \Big].
\end{equation}

\vskip 0.2cm
\noindent
The determinant of this metric can be written as follows:
\begin{equation}
\sqrt{- \mathrm{det} g} = \frac{\big(r^2 H^{\frac{1}{3}}\big)\, r
  \sin \theta \cos \theta }{\Xi_a \Xi_b}.
\end{equation}

\vskip 0.2cm
\noindent
Using this determinant, it is a long but straightforward calculation
to find the inverse metric. We find:
\begin{eqnarray}
r^2 H^{\frac{1}{3}} g^{t t} &=& - \frac{1}{4 g^6} \Bigg[ 12 g^4
  \Delta_{\theta} + \frac{4 \Xi_a \Xi_b \, g^4}{\Delta_{\theta}} +
  \frac{4 (\Xi_a \Xi_b)^{3/2} \mathcal{J}}{r^4} - \frac{ 4 \beta_2
  g^4}{r^4}  - \frac{4 \Xi_a \Xi_b \, g^2}{r^2} \nonumber \\
&+& \frac{1}{r^2 \Delta_r} \Big( - g^8 \beta_2^2 + 2 g^6 \beta_2 (2
r^2 + a^2 + b^2 + a^2 b^2 g^2) \nonumber \\
&-& (3 + 4 g^2 r_m^2 - \Xi_a \Xi_b)^2  \Big)  \Bigg], \nonumber \\
r^2 H^{\frac{1}{3}} g^{t \phi} &=& - \frac{\Xi_a}{2 g} \Bigg[ -
  \frac{2 \Xi_b}{\Delta_{\theta}} + \frac{ 2 \Xi_b r^2  + a^2 \Xi_b -
  4 b^2 g^2 r_m^2 - 2 b^4 g^2 - b^2 - \beta_2 g^2}{\Delta_r} \Bigg],
  \nonumber \\
r^2 H^{\frac{1}{3}} g^{t \psi} &=& - \frac{\Xi_b}{2 g} \Bigg[ -
  \frac{2 \Xi_a}{\Delta_{\theta}} + \frac{ 2 \Xi_a r^2  + b^2 \Xi_a -
  4 a^2 g^2 r_m^2 - 2 a^4 g^2 - a^2 - \beta_2 g^2}{\Delta_r} \Bigg],
  \nonumber \\
r^2 H^{\frac{1}{3}} g^{\phi \psi} &=&  \Xi_a \Xi_b
  \Bigg[\frac{r^2}{\Delta_r} -  \frac{1}{\Delta_{\theta}} \Bigg],
  \nonumber \\
r^2 H^{\frac{1}{3}} g^{\phi \phi} &=&  \Xi_a^2  \Bigg[\frac{\cot^2
    \theta}{\Delta_{\theta}}  +  \frac{r^2}{\Delta_r} \Bigg],
  \nonumber \\
r^2 H^{\frac{1}{3}} g^{\psi \psi} &=&  \Xi_b^2  \Bigg[\frac{\tan^2
    \theta}{\Delta_{\theta}}  +  \frac{r^2}{\Delta_r} \Bigg],
  \nonumber \\
r^2 H^{\frac{1}{3}} g^{r r} &=& \Delta_r, \nonumber \\ 
r^2 H^{\frac{1}{3}} g^{\theta \theta} &=& \Delta_{\theta}. 
\end{eqnarray}
This shows that $r^2 H^{\frac{1}{3}} \, g^{\mu \nu}$ is additively
separable as a function of $r$ and $\theta$. 

\vskip 0.2cm
\noindent
The constant $\mathcal{J}$ used above in the $g^{t t}$ component is
defined by the equation: 
\begin{equation}
\mathcal{J} = (1 + g^2 \mu_1)(1 + g^2 \mu_2)(1 + g^2 \mu_3).
\end{equation}

\subsection{The Hamilton-Jacobi equation} 
As in the previous two examples, the Hamilton-Jacobi equation
(\ref{hjeqn}) is only completely separable\footnote{That is, we can
  write the $F(r, \theta)$ part of (\ref{hjansatz}) as $F(r, \theta) =
  S_{r}(r) + S_{\theta}(\theta)$.}, for arbitrary charges and rotation
parameters if 
$m^2=0$. This is due to the fact that $H^{\frac{1}{3}}$ is not a 
simple function of $r$ and $\theta$.

\vskip 0.2cm
\noindent
If $m^2 = 0$, then the $\theta$-separation equation is
 \begin{eqnarray}
K &=& - \frac{1}{4 g^6} \Bigg[ 12 g^4
  \Delta_{\theta} + \frac{4 \Xi_a \Xi_b \, g^4}{\Delta_{\theta}}\Bigg]
  E^2 - \frac{2 \Xi_a \Xi_b}{ g \Delta_{\theta}}  E(\Phi +
  \Psi)\nonumber \\
&+& \frac{\Big( \Xi_a \cot \theta \, \Phi - \Xi_b \tan \theta \, \Psi
  \Big)^2 }{\Delta_{\theta}} + \Delta_{\theta}  \Bigg( \frac{ d
  S_{\theta}}{d \theta} \Bigg)^2 .
\end{eqnarray}
From the $\theta$ separation equation, we can extract a second order
tensor, $K^{\mu \nu}$, defined by $K = K^{\mu \nu} p_{\mu} p_{\nu}$
where the $p_{\mu}$ are the canonical momenta. If this complete
separation was valid for all $m^2$, then $K^{\mu \nu}$ would be a
Killing tensor, and would satisfy $ \nabla_{(\mu} K_{\nu \sigma)}
=0$. However, as the Hamilton-Jacobi equation only appears to separate
if $m^2 =0$, then we would only expect $K^{\mu \nu}$ to be a conformal
Killing tensor, which would satisfy  $ \nabla_{(\mu} K_{\nu \sigma)} =
g_{(\mu \nu} V_{\sigma)}$ for some co-vector field $V_{\sigma}$.

\vskip 0.2cm
\noindent
We find that $K^{\mu \nu}$ is given by:
\begin{eqnarray}
K^{\mu \nu} &=& - \frac{1}{4 g^6} \Bigg[ 12 g^4
  \Delta_{\theta} + \frac{4 \Xi_a \Xi_b \, g^4}{\Delta_{\theta}} \Bigg]
  \delta_t^{\mu} \delta_t^{\nu}  + \frac{ \Xi_a \Xi_b}{ g
  \Delta_{\theta}} (\delta_t^{\mu} \delta_{\phi}^{\nu} +
  \delta_t^{\nu} \delta_{\phi}^{\mu} + \delta_t^{\mu} \delta_{\psi}^{\nu} +
  \delta_t^{\nu} \delta_{\psi}^{\mu}) \nonumber \\
&+& \frac{\Xi_a^2 \cot^2 \theta}{\Delta_{\theta}} \delta_{\phi}^{\mu}
  \delta_{\phi}^{\nu}  + \frac{\Xi_b^2 \tan^2 \theta}{\Delta_{\theta}}
  \delta_{\psi}^{\mu} \delta_{\psi}^{\nu} - \frac{\Xi_a
  \Xi_b}{\Delta_{\theta}} (\delta_{\phi}^{\mu} \delta_{\psi}^{\nu} +
  \delta_{\phi}^{\nu} \delta_{\psi}^{\mu} ) + \Delta_{\theta}
  \delta_{\theta}^{\mu}   \delta_{\theta}^{\nu}  .
\end{eqnarray}
In order to find $V_{\sigma}$, we can use equation (\ref{cktvec}) with
the result:
\begin{equation}
V^{\mu} = \frac{\partial_{\theta}  H^{\frac{1}{3}}}{H^{\frac{1}{3}}}
  \Delta_{\theta} \, \delta_{\theta}^{\mu}
  \Longleftrightarrow V_{\mu} =  \partial_{\theta} ( r^2
  H^{\frac{1}{3}})   \, \delta^{\theta}_{\mu}. 
\end{equation}


\subsection{The Klein-Gordon equation}
As in the previous cases, we are unable to separate the Klein-Gordon
equation in the massive case due to the nature of the function
$H^{\frac{1}{3}}$. Taking the massless Klein-Gordon equation
(\ref{masslesskg}) and substituting into it the separable Ansatz
(\ref{kgansatz}), we find the following equation for $\Theta(\theta)$:
\begin{equation}
\frac{1}{\Theta \sin \theta \cos \theta} \frac{d}{d \theta} \Bigg(
\sin \theta \cos \theta \Delta_{\theta} \frac{d \Theta}{d \theta}
\Bigg) + \Bigg[\frac{\omega^2}{4 g^2}\Bigg(\frac{\Xi_a
    \Xi_b}{\Delta_{\theta}} + 3 \Delta_{\theta} \Bigg)+   \frac{2
    (\alpha + \beta) \omega \Xi_a \Xi_b}{g \Delta_{\theta}} \nonumber
\end{equation}
\begin{equation}
  - \frac{\alpha^2 \Xi_a^2  \cot^2 \theta}{\Delta_{\theta}} +
 \frac{2 \alpha \beta \Xi_a \Xi_b }{\Delta_{\theta}} -
 \frac{\beta^2 \Xi_b^2 \tan^2 \theta}{\Delta_{\theta}}
 \Bigg] = k,
\end{equation}
where $k$ is the separation constant. The analytic structure of this
equation can be seen more easily if we make the standard change of
variable $z=\sin^2 \theta$. This gives us  
\begin{eqnarray}
0 &=& \frac{d^2 \Theta}{d z^2} + \Bigg( \frac{1}{z} + \frac{1}{z-1} +
\frac{1}{z-d} \Bigg)\frac{d \Theta}{d z} + \Bigg[ \frac{\omega^2}{4
  g^2} \Bigg( \frac{\Xi_a \Xi_b}{z (1-z) \Delta_z^2} + \frac{3}{z
  (1-z)} \Bigg)\nonumber \\ 
&+& \frac{(\alpha + \beta) \omega \Xi_a \Xi_b}{2 g z (1-z) \Delta_z^2} -
\frac{\alpha^2 \Xi_a^2}{4 z^2 \Delta_z^2} - \frac{\beta^2 \Xi_b^2}{4
  (1 - z)^2 \Delta_z^2} + \frac{ \alpha \beta \Xi_a \Xi_b}{2 z (1-z)
  \Delta_z^2} - \frac{k}{4 z (1-z) \Delta_z} \Bigg] \Theta,
\end{eqnarray}
where $d= \Xi_a/g^2 (b^2 -a^2)$ and $\Delta_z = g^2 (a^2 -
b^2)(z-d)$. This differential equation has four regular singular
points and can be put in the form of Heun's equation. In the case
where the rotation parameters are equal, this differential equation
has regular singular points at $z=0$, $z=1$ and $z=\infty$ and the
solution can again be given in terms of hypergeometric functions. 


\section{General Results}

\noindent
A rank 2, conformal Killing tensor, $K^{\mu \nu}$ in $D$ dimensions is
a symmetric tensor that generates a non-trivial symmetry on the phase
space for massless particles.  From the perspective of integrability,
a non-trivial conformal Killing tensor gives rise to an extra constant
of the motion that is quadratic in the canonical momenta. Together
with the constants of motion due to the Killing vectors and conformal
Killing vectors of the spacetime, there are enough constants of the
motion to completely describe the geodesic motion of massless
particles giving rise to an integrable system. Conformal Killing
tensors obey the equation 
\begin{equation} \label{ckt2}
\nabla_{(\mu} K_{\nu \rho)} = g_{(\mu \nu} V_{\rho)},
\end{equation}  
for some co-vector field $V_{\rho}$. If $V_{\rho}=0$, then $K^{\mu
  \nu}$ is a Killing tensor. By contracting (\ref{ckt2}) with
$g^{\mu \nu}$, one obtains an equation for the associated vector
field, $V^{\mu}$:
\begin{equation}
V^{\mu} = \frac{2}{D+2} \nabla_{\nu} K^{\mu \nu} + \frac{1}{D+2}
\nabla^{\mu} K^{\nu}_{\phantom{\nu} \nu}. 
\end{equation} 

\vskip 0.2cm
\noindent
Clearly, any rank two tensor of the form $K^{\mu \nu} =
\Lambda(x^{\rho}) g^{\mu   \nu}$ will obey (\ref{ckt2}) with $V_{\mu} =
\nabla_{\mu} \Lambda$, so in that sense, any spacetime possesses an
infinite family of conformal Killing tensors just as any constant
multiplying the inverse spacetime metric will be a Killing
tensor. However, in certain circumstances, one finds second rank
tensors that obey (\ref{ckt2}), but which are not conformal to the
spacetime (inverse) metric. These are non-trivial conformal Killing
tensors. Given a non-trivial conformal Killing tensor, one can define
an equivalence class so that two conformal Killing tensors are in the
same equivalence class if they differ by some function multiplying the
inverse metric, that is:  
\begin{equation}
K^{\mu \nu} \sim L^{\mu \nu} \Longleftrightarrow K^{\mu \nu} = 
L^{\mu \nu} + \Lambda(x^{\rho}) g^{\mu \nu}
\end{equation}
for some function, $\Lambda$. If $V_{\mu}$ and $U_{\mu}$ are
the co-vectors associated to $K^{\mu \nu}$ and $L^{\mu \nu}$
respectively, then
\begin{equation}
V_{\mu} = U_{\mu} + \nabla_{\mu} \Lambda.
\end{equation}  
So, the associated vector fields transform within each equivalence
class like $U(1)$ gauge fields which implies that we can define a
gauge-invariant field strength on each equivalence class of conformal
Killing tensors: 
\begin{equation}
F_{\mu \nu} = \nabla_{\mu}
V_{\nu} - \nabla_{\nu} V_{\mu}.
\end{equation}  

\vskip 0.2cm
\noindent
We shall now prove a lemma that shows, under reasonable assumptions,
that cohomogeneity 2 metrics where the inverse metric can be
additively separated after multiplying through by some function admit
a conformal Killing tensor, and we show that the associated
(co-)vector takes the expected form. We shall prove this result in
five-dimensions, but generalisations to higher dimensions follow
trivially.

\vskip 0.2cm
\noindent
\emph{Lemma}: Let $g_{\mu \nu}$ be the metric on a five-dimensional
spacetime with coordinates $t, r, \theta, \phi$ and $\psi$ and let the
metric have cohomogeneity 2 so that $g_{\mu \nu} \equiv g_{\mu \nu}(r,
\theta)$. We shall also assume that the $(r, \theta)$ part of the
metric can be put into the diagonal form: 
\begin{equation}
d s^2_{r,\theta} = \Omega(r, \theta) \Big( \Theta^2 (\theta) d
\theta^2 +  R^2(r) d r^2 \Big),
\end{equation} 
and suppose that $\Omega (r, \theta) \, g^{\mu \nu}$ can be written as a
tensor depending only on $r$ plus a tensor depending only on
$\theta$\footnote{The first two of the solutions that we studied can
  be put into the canonical form given in \cite{cclp3} which possesses
  all of these properties. The third solution that we studied
  \cite{klr} also matches these criteria. Therefore, these assumptions
  on the properties of the metric are reasonable.}: 
\begin{equation} \label{Pdef}
\Omega (r, \theta) g^{\mu \nu} = P^{\mu \nu} (\theta) + S^{\mu \nu}(r).
\end{equation}
Then, $P^{\mu \nu}(\theta)$ and $S^{\mu \nu}(r)$ are conformal Killing tensors
with corresponding vector fields
\begin{equation}
V^{\mu} = (\partial_{\theta} \Omega) g^{\theta \theta}
\delta_{\theta}^{\mu} 
\end{equation}
and
\begin{equation}
U^{\mu} = (\partial_{r} \Omega) g^{r r}\delta_{r}^{\mu}
\end{equation}
respectively. 

\vskip 0.2cm
\noindent
\emph{Proof}: Let $P^{\mu \nu}(\theta)$ be defined by equation
(\ref{Pdef}). We shall evaluate (\ref{ckt2}) with all three indices
raised in order to minimise calculational effort. 
\begin{eqnarray}
\nabla^{(\mu} P^{\nu \rho)} &=& g^{\sigma (\mu}  \partial_{\sigma}
P^{\nu \rho)} + g^{\sigma (\mu} \Gamma_{\sigma \tau}^{\nu} P^{\rho)
  \tau} + g^{\sigma (\mu} \Gamma_{\sigma \tau}^{\rho} P^{\nu) \tau} 
\nonumber \\
&=& g^{\theta (\mu} \partial_{\theta} P^{\nu \rho)} + \frac{1}{2}
g^{\sigma (\mu} g^{\nu |\alpha|} (g_{\sigma \alpha, \tau} + g_{\tau
  \alpha, \sigma} - g_{\sigma \tau, \alpha}) P^{\rho) \tau} 
\nonumber \\
&+& \frac{1}{2}
g^{\sigma (\mu} g^{\rho |\alpha|} (g_{\sigma \alpha, \tau} + g_{\tau
  \alpha, \sigma} - g_{\sigma \tau, \alpha}) P^{\nu) \tau}
\nonumber \\
&=& g^{\theta \theta} \delta^{(\mu}_{\theta}\partial_{\theta} \Big(
  \Omega g^{\nu\rho)} \Big) + \frac{1}{2}  g^{\sigma (\mu} g^{\nu
  |\alpha|} g_{\sigma \alpha, \theta} P^{\rho) \theta} + \frac{1}{2}
  g^{\sigma (\mu} g^{\rho
  |\alpha|} g_{\sigma \alpha, \theta} P^{\nu) \theta}
\nonumber \\
&=& g^{\theta \theta} \delta^{(\mu}_{\theta} g^{\nu \rho)}
\partial_{\theta} \Omega  + \Omega g^{\theta \theta} 
\delta^{(\mu}_{\theta} \partial_{\theta} g^{\nu \rho)}  + 
\Omega g^{\theta \theta} \delta^{(\mu}_{\theta}   g^{\sigma (\mu} g^{\nu
  |\alpha|} \delta^{\rho)}_{\theta} g_{\sigma \alpha, \theta} 
\nonumber \\
&=& g^{(\mu \nu} \Big( \delta_{\theta}^{\rho)} g^{\theta \theta}
\partial_{\theta} \Omega \Big). 
\end{eqnarray}
In performing this calculation, we have made use of the facts that
$P^{r \mu} = 0$, $P^{\mu \theta} = P^{\theta \theta}
\delta_{\theta}^{\mu}$ as the $(r,\theta)$ part of the metric is
diagonal and $\partial_{\mu} P^{\nu \rho} = \delta^{\theta}_{\mu}
\partial_{\theta} P^{\nu \rho}$ as $P^{\mu \nu}$ is a function of
$\theta$ only. The $|\alpha|$ notation means that the index $\alpha$
should not be included in the symmetrisation of the indices. 

\vskip 0.3cm
\noindent
Thus, we have proved that $P^{\mu \nu}(\theta)$ is a conformal
Killing tensor. Moreover, we have shown that the associated vector
field is $V^{\mu} = \delta_{\theta}^{\mu} g^{\theta \theta}
\partial_{\theta} \Omega$ and the co-vector field is $V_{\mu} =
\delta^{\theta}_ {\mu} \partial_{\theta} \Omega$. The corresponding
result from the $r$-dependent conformal Killing tensor, $S^{\mu \nu}$
is $U_{\mu} =  \delta^{r}_{\mu}  \partial_r \Omega$. 

\vskip 0.2cm
\noindent
It is immediately apparent that the two conformal Killing tensors,
$P^{\mu \nu}$ and $S^{\mu \nu}$, have different vector fields
associated to them. However, up to a sign, they give rise to the same
constant of the motion when one separates the Hamilton-Jacobi
equation. To see this, we consider the Hamilton-Jacobi equation, 
\begin{equation} \label{HJ}
\frac{ \partial S}{\partial \tau} + \frac{1}{2} \, g^{\mu \nu} \frac{
  \partial S}{\partial x^{\mu}} \frac{ \partial S}{\partial x^{\nu}} = 0
\end{equation}
and we seek a separable solution to it, of the form\footnote{The black
  hole spacetimes that we have considered have Killing vectors
  $\partial_t, \partial_{\phi}$ and $\partial_{\psi}$, so these
  coordinates appear linearly in Hamilton's principal function.}
\begin{equation}
S = \frac{1}{2} m^2 \tau - E t + \Phi \phi + \Psi \psi + S_{r}(r) +
S_{\theta}(\theta). 
\end{equation}
Inserting this into (\ref{HJ}), and multiplying through by $2\Omega$,
we find:
\begin{eqnarray}
0 &=& m^2 \Omega + (-E)^2 (P^{t t} + S^{t t}) + 2 (-E) \Phi
(P^{t \phi} + S^{t \phi}) + 2 (-E) \Psi (P^{t \psi} + S^{t \psi}) 
\nonumber \\
&+& \Phi^2 (P^{\phi \phi} + S^{\phi \phi}) + 2 \Psi \Phi (P^{\phi \psi}
+ S^{\phi \psi}) + \Psi^2 (P^{\psi \psi} + S^{\psi \psi}) \nonumber \\
&+& P^{\theta
  \theta} \Bigg( \frac{ d S_{\theta}}{d \theta}
\Bigg)^2 + S^{r r} \Bigg( \frac{ d S_{r}}{d r}
\Bigg)^2. 
\end{eqnarray}
The function $\Omega$ is not generally the sum of a function of $r$
and a function of $\theta$, so the $r$ and $\theta$ dependence can
only be separated in the massless case, that is if $m^2 =0$. Given a
separable solution to the Hamilton-Jacobi equation, we can deduce the
geodesic equations up to quadratures. Having set $m^2 = 0$, the $r$ and 
$\theta$ dependence can be separated as follows:
\begin{eqnarray}
K &=&  (-E)^2 P^{t t}  + 2 (-E) \Phi P^{t \phi} + 2 (-E) \Psi P^{t \psi} 
\nonumber \\
&+& \Phi^2 P^{\phi \phi} + 2 \Psi \Phi P^{\phi \psi} + \Psi^2 P^{\psi
  \psi}   + P^{\theta
  \theta} \Bigg( \frac{ d S_{\theta}}{d \theta}
\Bigg)^2, 
\end{eqnarray}
and
\begin{eqnarray}
-K &=&  (-E)^2  S^{t t} + 2 (-E) \Phi S^{t \phi} + 2 (-E) \Psi S^{t \psi} 
\nonumber \\
&+& \Phi^2 S^{\phi \phi} + 2 \Psi \Phi S^{\phi \psi} + \Psi^2  S^{\psi
  \psi} + S^{r r}  \Bigg( \frac{ d S_{r}}{d r}
\Bigg)^2. 
\end{eqnarray}
Letting $K = K^{\mu \nu} p_{\mu} p_{\nu}$, where $K^{\mu \nu}$ is the
conformal Killing tensor, we have a choice. We can choose $K^{\mu \nu}
= P^{\mu \nu}$ or $K^{\mu \nu} = - S^{\mu \nu}$. So, the two possible
conformal Killing tensors differ by $\Omega g^{\mu \nu}$, and thus lie
in the same equivalence class. This also explains the difference in
the associated vector fields. 

\vskip 0.2cm
\noindent
Independently of the choice of conformal Killing tensor, the only
non-zero components of the associated field strength, $F_{\mu \nu}$,
are $F_{r \theta} = - F_{\theta r} = \partial_r  \partial_{\theta}
\Omega$. In the case where $\Omega = f(r) + g(\theta)$, this field
strength vanishes which is unsurprising because in that case, we would
have a Killing tensor and not a conformal Killing tensor.


\subsection{Relation to Killing tensors}

Here we consider the case when $\Omega(r,\theta)$ can be written as a
function of $r$ plus a function of $\theta$. This will give rise to
Killing tensors rather than conformal Killing tensors. Let
$\Omega(r,\theta) = f(\theta) + h(r)$. Then the equation for the
separation of the inverse metric can be written as:
\begin{equation}
(f(\theta) + h(r)) g^{\mu \nu} = P^{\mu \nu}(\theta) + S^{\mu
    \nu}(r). 
\end{equation}
As it stands, $P^{\mu \nu}$ and $S^{\mu \nu}$ are conformal
Killing tensors, but it is easy to show that 
\begin{equation}
P'^{\mu \nu} = P^{\mu \nu}(\theta) - f(\theta) g^{\mu \nu} \quad
\mathrm{and} \quad S'^{\mu \nu} = S^{\mu \nu}(r) - h(r) g^{\mu \nu}  
\end{equation}  
are Killing tensors. In fact, it is very easy to see that $ P'^{\mu
  \nu} = - S'^{\mu \nu}$.

\subsection{Relevance of the vector field}
We have seen that the constant of motion that arises in the separation
of the massless Hamilton-Jacobi equation can be expressed in terms of
a conformal Killing tensor, and that in reality, there are two
possible conformal Killing tensors that will give rise to the same
constant of motion. As we discussed earlier, these two conformal
Killing tensors have different vector fields associated to them, so
this raises the question as to what, if any, physical relevance can be
attached to these vector fields. 

\vskip 0.2cm
\noindent
If we compute the Poisson bracket of $K = K^{\mu \nu} p_{\mu} p_{\nu}$ with 
the Hamiltonian, $H = g^{\mu \nu} p_{\mu} p_{\nu}$, we find the rather
simple result
\begin{equation}
\{ K,H \} = 2 H (V^{\mu} p_{\mu}),
\end{equation}
where $V^{\mu}$ is the vector field associated to the conformal
Killing tensor, $K^{\mu \nu}$, and $p_{\mu}$ are the canonical
momenta.  

\vskip 0.2cm
\noindent
The Poisson bracket of a function with the Hamiltonian describes the
evolution of the function along the particle worldline:
\begin{equation}
\frac{d f}{d \tau} = \{ f,H \},
\end{equation}
so we see that on massless geodesics (where $H =0$) $K$ is constant as
expected. We also see the importance of the vector field $V^{\mu}$. On
massive paths, it controls the evolution of the scalar quantity $K$ as
you go along the particle worldline. Performing the necessary
integration, we obtain
\begin{equation} \label{K}
K(\tau) = K_0 - m^2 \int_0^{\tau} V^{\mu} p_{\mu} \, d \tau'.
\end{equation}
The scalar $K$ differs from the scalar obtained from the other
conformal Killing tensor by $m^2 \Omega(\tau)$ in the massive case. 


\section{Conclusions}
In this paper, we have studied three black hole solutions of five dimensional
supergravity with three charges and independent rotation
parameters. In each case, we found that the inverse metric multiplied
by some function can be separated, additively, as a function of the
radial variable, $r$, and an angular variable, $\theta$. However, the
function that we multiply the inverse metric by takes a rather
unpleasant form unless we simplify matters by taking equal rotation
parameters or equal charges. This means that the Hamilton-Jacobi
equation which governs geodesic flow on the cotangent bundle cannot be
separated apart from in the case of massless particles. The
separability in the massless case is linked to the existence of a rank
two, conformal Killing tensor which we found in all cases. A rank two,
conformal Killing tensor obeys an equation that involves a
(co-)vector field. For the black hole solutions under investigation
here, we have computed these (co-)vector fields and their rather
simple expressions have allowed us to suggest and prove a general
form. 

\vskip 0.2cm
\noindent
The separation of the Hamilton-Jacobi equation involves the
introduction of a separation constant which appears in both the $r$
and $\theta$ separation equations. Therefore, one can extract two
conformal Killing tensors for each black hole, although they differ by
a function multiplying the inverse metric. The associated (co-)vector
fields then differ by the gradient of the aforementioned function. The
physical relevance of these (co-)vectors can be seen when considering
the evolution of the quantity $K^{\mu \nu} p_{\mu} p_{\nu}$ along a
massive particle worldline. 

\vskip 0.2cm
\noindent
Finally, we have showed that the Klein-Gordon equation is also
separable in the massless case for all three black holes considered
here. 


\section{Acknowledgements}
The author would like to thank Gary Gibbons, Hari K. Kunduri, James
Lucietti and Chris Pope for many useful discussions and comments on
this manuscript.



\thebibliography{99}

\bibitem{maldacena}
J.~M.~Maldacena, ``The large N limit of superconformal field theories
and supergravity,'' Adv.\ Theor.\ Math.\ Phys.\ {\bf 2} (1998) 231,
{\tt hep-th/9711200}.
 
\bibitem{mp}
R.~C.~Myers and M.~J.~Perry, ``Black holes in higher dimensional
space-times,'' Annals Phys.\ {\bf 172} (1986) 304.

\bibitem{glpp}
G.~W.~Gibbons, H.~L\"u, D.~N.~Page and C.~N.~Pope, ``The general
Kerr-de Sitter metrics in all dimensions,'' J.\ Geom.\ Phys.\ {\bf 53}
(2005) 49, {\tt hep-th/0404008}.

\bibitem{stoj}
V.~P.~Frolov and D.~Stojkovic, ``Particle and light motion in a
space-time of a five-dimensional rotating black hole,'' Phys.\ Rev.\ D
{\bf 68} (2003) 064011, {\tt gr-qc/0301016}.

\bibitem{kl1}
H.~K.~Kunduri and J.~Lucietti, ``Integrability and the Kerr-(A)dS
black hole in five dimensions,'' Phys.\ Rev.\ D {\bf 71} (2005) 104021,
{\tt hep-th/0502124}.

\bibitem{vs}
M.~Vasudevan and K.~A.~Stevens, ``Integrability of particle motion and
scalar field propagation in Kerr-(Anti) de Sitter black hole
spacetimes in all dimensions,'' Phys.\ Rev.\ D {\bf 72} (2005) 124008,
{\tt gr-qc/0507096}.

\bibitem{clp4}
M.~Cveti\v c, H.~L\"u and C.~N.~Pope, ``Charged Kerr-de Sitter black
holes in five dimensions,'' Phys.\ Lett.\ B {\bf 598} (2004) 273, {\tt
  hep-th/0406196}.

\bibitem{kl2}
H.~K.~Kunduri and J.~Lucietti, ``Notes on non-extremal, charged,
rotating black holes in minimal $D=5$ gauged supergravity,'' Nucl.\
Phys.\ B {\bf 724} (2005) 343, {\tt hep-th/0504158}.

\bibitem{cclp2}
Z.~W.~Chong, M.~Cveti\v c, H.~L\"u and C.~N.~Pope, ``General
non-extremal rotating black holes in minimal five-dimensional gauged
supergravity,'' Phys.\ Rev.\ Lett.\ {\bf 95} (2005) 161301, {\tt
  hep-th/0506029}. 

\bibitem{dkl}
P.~Davis, H.~K.~Kunduri and J.~Lucietti, ``Special symmetries of the
charged Kerr-AdS black hole of $D=5$ minimal gauged supergravity,''
Phys.\ Lett.\ B {\bf 628} (2005) 275, {\tt hep-th/0508169}.

\bibitem{klemm}
D.~Klemm, ``Rotating black branes wrapped on Einstein spaces,'' JHEP\
{\bf 9811} (1998) 019, {\tt  hep-th/9811126}.

\bibitem{cglp}
Z.~W.~Chong, G.~W.~Gibbons, H.~L\"u and C.~N.~Pope, ``Separability and
Killing tensors in Kerr-Taub-NUT-de Sitter metrics in higher
dimensions,'' Phys.\ Lett.\ B {\bf 609} (2005) 124, {\tt
  hep-th/0405061}. 

\bibitem{clp}
W.~Chen, H.~L\"u and C.~N.~Pope, ``Kerr-de Sitter black holes with NUT
charges,'' to appear in Nucl.\ Phys.\ B, {\tt hep-th/0601002}.

\bibitem{clp2}
W.~Chen, H.~L\"u and C.~N.~Pope, ``Separability in cohomogeneity-2
Kerr-NUT-AdS metrics,'' JHEP {\bf 04} (2006) 008, {\tt hep-th/0602084}.

\bibitem{davis}
P.~Davis, ``A Killing tensor for higher dimensional Kerr-AdS black
holes with NUT charge,'' Class.\ Quantum\ Grav.\ {\bf 23} (2006) 3607,
{\tt hep-th/0602118}.

\bibitem{clp3}
W.~Chen, H.~L\"u and C.~N.~Pope, ``General Kerr-NUT-AdS metrics in all
dimensions,''{\tt hep-th/0604125}.

\bibitem{cy}
M.~Cveti\v c and D.~Youm, ``General rotating five dimensional black
holes of toroidally compactified heterotic string,'' Nucl.\ Phys.\ B
{\bf 476} (1996) 118, {\tt hep-th/9603100}.

\bibitem{cclp}
Z.~W.~Chong, M.~Cveti\v c, H.~L\"u and C.~N.~Pope, ``Five-dimensional
gauged supergravity black holes with independent rotation
parameters,'' Phys.\ Rev.\ D {\bf 72} (2005) 041901, {\tt
  hep-th/0505112}. 

\bibitem{klr}
H.~K.~Kunduri, J.~Lucietti and H.~S.~Reall, ``Supersymmetric
multi-charge $AdS_5$ black holes,'' JHEP {\bf 04} (2006) 036, {\tt
  hep-th/0601156}.

\bibitem{cclp3}
Z.~W.~Chong, M.~Cveti\v c, H.~L\"u and C.~N.~Pope, ``Non-extremal
rotating black holes in five-dimensional gauged supergravity,'' {\tt
  hep-th/0606213}. 

\bibitem{bmpv}
J.~C.~Breckenridge, R.~C.~Myers, A.~W.~Peet and C.~Vafa, ``D-branes
and spinning black holes,'' Phys.\ Lett.\ B {\bf 391} (1997) 93, {\tt
  hep-th/9602065}.

\bibitem{vsp}
M.~Vasudevan, K.~A.~Stevens and D.~N.~Page, ``Separability of the
Hamilton-Jacobi and Klein-Gordon equations in Kerr-de Sitter
metrics,'' Class.\ Quant.\ Grav.\ {\bf 22} (2005) 339, {\tt
  gr-qc/0405125}.

\bibitem{carter}
B.~Carter, ``Hamilton-Jacobi and Schrodinger separable solutions of
Einstein's equations,'' Comm.\ Math.\ Phys.\ {\bf 10} (1968) 280.
\end{document}